\documentclass{article}
\usepackage{graphicx} 
\usepackage{comment}
\usepackage[utf8]{inputenc}
\usepackage{amsmath, amssymb}
\usepackage{authblk}
\usepackage{textcomp}
\usepackage{float}
\usepackage[colorlinks=true, linkcolor = black, urlcolor=blue,citecolor=blue]{hyperref}

\usepackage[labelfont=bf]{caption} 

\usepackage{geometry}
\usepackage{url}         
\usepackage{authblk}

\title{Mid-infrared integrated resonators on a III-V platform with Q-factors beyond half a million}
\author[1]{Luca Lucia}
\author[1]{Gia Long Ngo}
\author[1]{Mathilde Urbain}
\author[1]{Konstantinos Pantzas}
\author[1]{Grégoire Beaudoin}
\author[1]{Stefano Pirotta}
\author[1]{Jean-Michel Manceau}
\author[1]{Isabelle Sagnes}
\author[1]{\newline Raffaele Colombelli}
\author[1]{Adel Bousseksou}

\affil[1]{Centre de Nanosciences et de Nanotechnologies (C2N), CNRS UMR 9001, Université Paris-
Saclay, Palaiseau 91120, France.
}

\date{}

\begin{document}

\maketitle

\begin{abstract}
We demonstrate mid-IR integrated race-track resonators on a III-V semiconductor platform: InGaAs core epitaxially grown on InP. We have performed a complete characterization of the optical propagation losses at $\lambda$ = 4.6~µm and $\lambda$ = 8.5~µm, two representative wavelengths for the $1^{\textnormal{st}}$ and $2^{\textnormal{nd}}$ atmospheric transparency windows. We measured losses of 1~dB/cm (TE polarization) and 1.3~dB/cm (TM polarization) at 8.5~µm. Substantially lower losses, 0.28~dB/cm, were measured at $\lambda$ = 4.6~µm (TM polarization). We then implemented racetrack resonators with straight evanescent couplers. We obtained loaded quality factors larger than 600.000 at a $\lambda$ = 4.6~µm. These results are promising towards the development of non-linear mid-IR integrated devices with Q factors beyond $10^{6}$, where the onset of stimulated parametric processes could be reachable.
\end{abstract}

\section{Introduction}

The invention and development of the frequency comb has opened a vast range of possibilities in several application domains~\cite{REF1_hansch2005nobel}. In the VIS/NIR (visible, near-infrared) spectral ranges, such technology has led to radical advances in metrology and spectroscopy \cite{REF2_udem2002optical}. In the mid-infrared (MIR, 3~µm $\leq$ $\lambda$ $\leq$ 30~µm) spectral range, dual-comb spectroscopy has become a particularly valuable technique. One major advantage is the possibility to acquire broadband spectra at much faster speeds than conventional FTIR spectrometers (hundreds of us instead of tens of seconds)~\cite{REF3_Picque2018}~\cite{REF4_Millot2016}.\\
Frequency combs are traditionally generated via mode-locked lasers. When the phases are all equal, an amplitude modulated comb is produced (TiSa, for instance)~\cite{REF5_MA2019}, and pulses are generated. If the phases are locked, but not equal, it is possible to obtain frequency modulated (FM) combs. FM combs are typically favored in lasers with upper state lifetimes shorter than the photon lifetime, such as quantum cascade lasers (QCL). As a matter of fact, QCL based FM combs have recently seen an intense development~\cite{REF6_FAIST2012}~\cite{REF7_FAIST2014}~\cite{REF8_Burghoff}~\cite{REF9_Piccardo2020}.\\
About 15 years ago, an alternative strategy to generate frequency combs was discovered, based on the Kerr effect in very high Q-resonators~\cite{REF10_Chembo}~\cite{REF11_DelHaye2007}. Now referred to as “Kerr combs”, it relies on the interaction between a continuous-wave pump laser with a monolithic ultra-high-Q resonator, and it exploits the onset of four-wave-mixing (FWM) gain under specific conditions. The discovery enabled the development of integrated micro-comb sources at telecom wavelengths~\cite{REF12_Chang2020}. With the current trend to develop photonic integrated circuits (PIC) in the MIR, pushed by spectroscopy and trace-gas sensing applications, it is natural to try and develop integrated micro-combs operating at long infrared wavelengths.\\
The minimum pump power threshold for parametric gain is inversely proportional to the resonator total Q-factor (\textnormal{$Q_{tot}$}) and to the nonlinear index of refraction (\textnormal{$n_{2}$})~\cite{REF10_Chembo} (among other parameters). For this reason, the community is devoting a vast, coordinated effort to explore, and optimize, material platforms for very low waveguide losses and elevated \textnormal{$n_{2}$}. To this day, losses in the (sub)-dB/cm range have been reported in Ge-on-Si~\cite{REF13_Armand2023}, graded Ge on Si~\cite{REF14_Turpaud}, and in III-V materials (GaAs/AlGaAs, InGaAs/InP~\cite{REF15_Zhang}~\cite{REF16_MontesinosBallester2024}). As a consequence, resonators with Q-factors of few hundred thousand have been reported too. The situation is different for \textnormal{$n_{2}$}, whose values for the different materials are not precisely known at long infrared wavelengths ($\lambda$~$>$~3~µm)~\cite{REF17_Ensley}~\cite{REF18_Zhang}. It is therefore still unclear what is the best material to implement Kerr dissipative combs in the MIR.\\
In this letter, we report mid-IR low-loss waveguides and racetrack resonators in Metalorganic vapor-phase epitaxy (MOVPE)-grown In\textnormal{$_{0.53}$}Ga\textnormal{$_{0.47}$}As-on-InP materials platform with measured ultra-high Q-factors exceeding 600.000 (corresponding to intrinsic Q$_{s}$ exceeding 700.000) at $\lambda$~=~4.6~µm. The result is corroborated by a full characterization of the optical propagation losses in this system at representative wavelengths in the two atmospheric windows (4.6~µm and 8.5~µm). Our findings suggest that the In\textnormal{$_{0.53}$}Ga\textnormal{$_{0.47}$}As-on-InP platform is a promising system to demonstrate mid-IR Kerr combs, with eventual full integration with active components  such as quantum cascade lasers and quantum cascade detectors.

\section{Waveguide loss}
The device wafer was grown in a MOVPE reactor on a Fe-doped InP substrate, acting also as waveguide bottom cladding. A 3.5-µm-thick InGaAs layer was grown as waveguide core on top of the Fe-InP substrate and a non-intentionally doped InP top layer acts as waveguide top cladding. 
\subsection{Waveguide design and fabrication}
\label{sec:Section_2a}
The design of the waveguides was performed to achieve single-mode operation for both TE and TM polarizations at $\lambda$ = 8.5~µm. Finite difference eigenmode (FDE) simulations  were used to obtain the mode effective indices and the final ridge width was fixed to 6~µm. Adiabatic tapers 20~µm-wide and 750~µm-long were designed using eigenmode expansion (EME) simulations for optimized mid-IR light injection and collection. The minimum bending radius was set to 700~µm to avoid radiation losses in the bent sections. To ease the experimental alignment and measurement procedures, the input-output facets of each spiral were offset by 3~mm. 

\begin{figure}[htbp]
  \centering
  \includegraphics[width=0.8\textwidth]{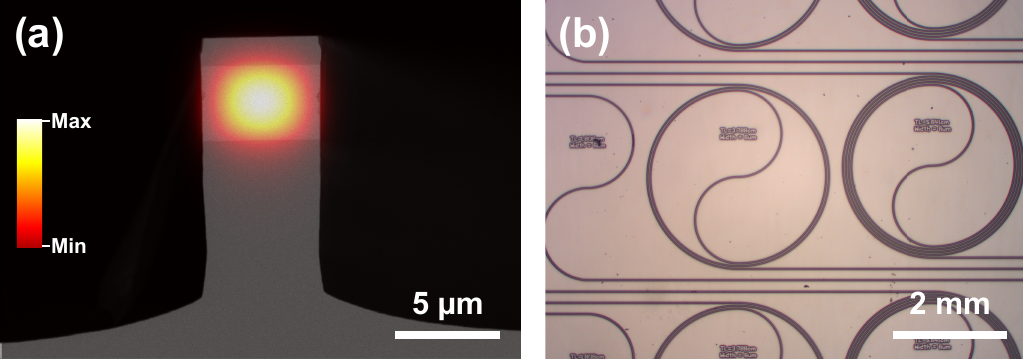} 
  \caption{\textbf{Profile and geometry of the spiral waveguides}. \textbf{(a)} SEM image of the etched waveguide profile, overlaid with the simulated electric field intensity profile of the fundamental TM\textnormal{$_{00}$} mode at 8.5~µm wavelength. \textbf{(b)} Optical microscope image (top view) of the spiral waveguides with varying lengths.}
  \label{fig:Fig1}
\end{figure}

To assess the performances of our InP platform, we performed loss measurements on dielectric waveguides using spiral waveguides of different lengths (1.8~cm, 3.8~cm, 5.6~cm, each repeated 3 times on the same chip). A top view optical microscope image of the final fabricated sample is shown in Fig.~\ref{fig:Fig1}(b). The sample fabrication starts with deposition of a Si\textnormal{$_{x}$}N\textnormal{$_{y}$} hard-mask by plasma-enhanced chemical vapor deposition (PECVD). Subsequently, the waveguides are defined with electron-beam lithography (EBL) and the pattern is transferred to the hard-mask with reactive ion etching (RIE). The semiconductor stack is etched using an inductively-coupled plasma RIE (ICP-RIE) based on HBr/O\textnormal{$_{2}$} chemistry. Finally, the Si\textnormal{$_{x}$}N\textnormal{$_{y}$} hard-mask is removed in pure hydrofluoric acid. The sample is then cleaved to obtain clean facets on both input and output. Fig.~\ref{fig:Fig1}(a) shows a scanning electron microscope (SEM) image of the waveguide cross section with the calculated electric field intensity profile of the fundamental TM\textnormal{$_{00}$} mode overlaid on it.

\subsection{Loss measurements and results}
\label{sec:Section_2b}
The waveguide loss measurements were performed using two continuous wave distributed feedback quantum cascade lasers (DFB-QCL) emitting at $\lambda$ = 8.5~µm and $\lambda$ = 4.6~µm, respectively. The emission wavelength was tuned by finely changing the laser driving current. The transmission spectrum of each waveguide was recorded using a liquid nitrogen-cooled mercury cadmium telluride (MCT) detector and a lock-in amplifier. 
The waveguide losses were inferred using the method described in \cite{REF19_Taebi}. The waveguide transmission as a function of the laser emission wavelength presents Fabry-Pérot (FP) oscillations stemming from the cavity induced by the cleaved facets. A typical FP spectrum measured around $\lambda$ = 8.5~µm is shown in the insets of Fig. 2(a) and 2(b) for the shortest spiral in both TM and TE polarization, respectively. The waveguide losses can be related to FP maxima, minima and facet reflectivity through Eq.~\ref{eq:Eq1}: 

\begin{equation}
\alpha = - \frac{1}{L} \ln\left( \frac{1}{R} \cdot \frac{\sqrt{r} - 1}{\sqrt{r} + 1} \right)
\label{eq:Eq1}
\end{equation}

where $r = \frac{I_{max}}{I_{min}}$ is the ratio between the maxima and minima of the FP oscillations; $R = \sqrt{R_{1}R_{2}}$ (with R\textnormal{$_{1}$} and R\textnormal{$_{2}$} being the reflectivity of each facet). An equivalent way to express Eq.~\ref{eq:Eq1} is:

\begin{equation}
\ln\left( \frac{1}{R} \cdot \frac{\sqrt{r} - 1}{\sqrt{r} + 1} \right) = - \alpha L - \ln \left( \frac{1}{R}\right)
\label{eq:Eq2}
\end{equation}

\begin{figure}[H]
  \centering
  \includegraphics[width=0.7\textwidth]{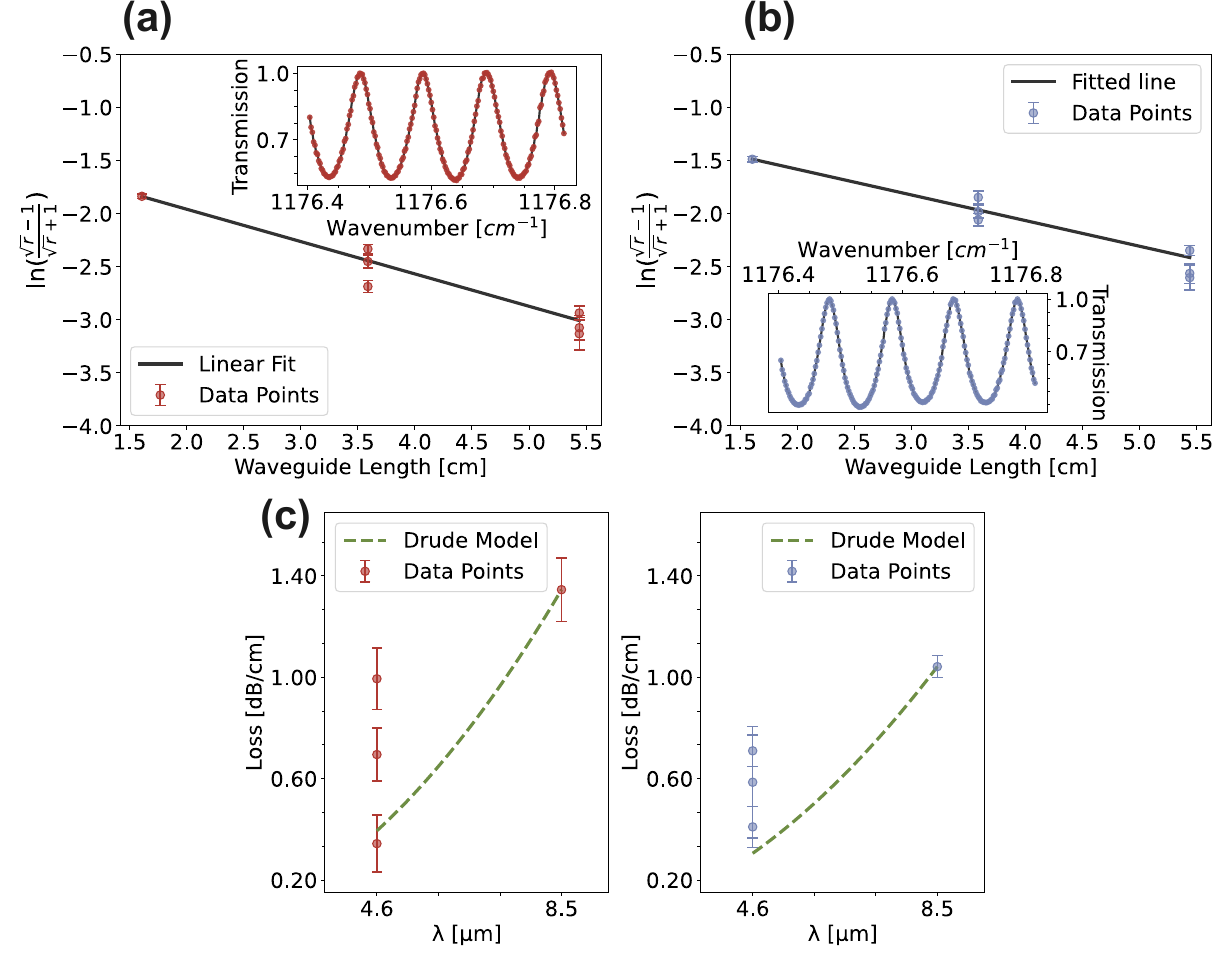} 
  \caption{\textbf{Propagation loss analysis of the fabricated waveguides.} \textbf{(a)} Plot of $\ln \left( \frac{\sqrt{r} - 1}{\sqrt{r} + 1}  \right)$ as a function of waveguide length for three different lengths in TM polarization. The linear fit (solid black line) yields $\alpha$~=~1.3$\pm$0.05~ dB/cm and R~=~0.26$\pm$0.02. The inset shows a typical transmission spectrum with Fabry–Pérot oscillations. \textbf{(b)} Same analysis as \textbf{(a)} for TE polarization, yielding $\alpha$~=~1.04$\pm$0.04~dB/cm and R~=~0.33$\pm$0.01. \textbf{(c)} Comparison of measured loss values at a wavelength of 8.5~µm and 4.6~µm for TM (left) and TE (right) polarization. The dashed curve represents the predicted loss change, based on the 8.5~µm loss value, considering free carrier absorption as main source of losses. Variability in the estimated loss value at 4.6~µm is ascribed to waveguide multimode operation. Very low losses, down to 0.3~dB/cm, are predicted at $\lambda$~=~4.6~µm.}
  \label{fig:Fig2}
\end{figure}

Having measured the FP oscillations for spirals with different lengths, the quantity $\ln \left( \frac{\left( \sqrt{r}-1\right)}{\left( \sqrt{r}+1\right)} \right)$ is known and can be plot as function of the waveguide length according to Eq.~\ref{eq:Eq2}. A linear curve is obtained from the plot, whose slope is $\alpha$, the loss per unit length. The intercept with y-axis yields the facet reflectivity. This method is more reliable than estimating $\alpha$ from a single FP transmission spectrum using Eq.~\ref{eq:Eq1}. The reason is that it \textit{directly} measures the facet reflectivity, and avoids assuming a value, usually calculated with numerical simulations. The mean and standard deviation of the fringe contrast for each spiral was measured at $\lambda$ = 8.5~µm and used to perform a linear fit with Eq.~\ref{eq:Eq2}. As result of a first run of measurements, and following an additional HF deoxydation, we obtained $\alpha$ = (1.3 $\pm$ 0.05)~dB/cm for TM and (1.04 $\pm$ 0.04)~dB/cm for TE polarization (see Fig. ~\ref{fig:Fig2}(a) and ~\ref{fig:Fig2}(b)). These results validate ICP-etched InP-based semiconductor platform as a valid low-loss system for mid-IR PICs. Note: judicious deoxidation steps play a crucial role in reducing the waveguide losses, especially for TE polarization, that exhibits the highest overlap with the waveguide sidewalls.

The same measurements were performed at $\lambda$ = 4.6~µm. However, because the waveguides were designed for single mode operation at 8.5~µm, they are multimode at shorter wavelengths. As a consequence, depending on the position of the laser beam on the input taper ridge, multiple mode excitation is possible. Whenever multiple modes are excited, we consistently saw smaller mean values of the fringe contrast with respect to single mode excitation (the multimode operation was clearly visible on the mid-IR camera when looking at the mode profile at the output facet of the waveguide). To account for this effect, different loss values were calculated with Eq.~\ref{eq:Eq1} for the shortest spiral at 4.6~µm. The loss values range from a minimum of 0.3~dB/cm (when the fringe contrast is the highest) to a maximum of 1~dB/cm, depending on the excitation condition. Also shown in the same graph is the value measured at 8.5~µm. Considering free carrier absorption as the primary source of losses in our platform, a simple Drude model can provide an estimate of the expected losses at 4.6~µm. The calculated trend (green dashed line in Fig. ~\ref{fig:Fig2}(c)) is approximately 0.3~dB/cm, aligning more closely with the measured minimum losses (corresponding to single mode operation) rather than the maximum.

An alternative way to gauge the losses is to estimate the intrinsic quality factor (\textnormal{$Q_{intrinsic}$}) of resonators. This quantity is directly linked to the propagation losses (losses per unit length) through Eq.~\ref{eq:Eq3}~\cite{REF19_Taebi}~\cite{REF15_Zhang}:

\begin{equation}
Q_{intrinsic} = \frac{2\pi n_{g}}{\alpha \lambda}
\label{eq:Eq3}
\end{equation}

where n$_{g}$ is the mode group index. Furthermore, losses as low as 0.3 dB/cm, at $\lambda$ = 4.6~µm, yield a Q$_{intrinsic}$ well beyond 500.000. To the best of our knowledge, the highest Q$_{intrinsic}$ measured on an InP platform in the mid-IR is 175.000 at $\lambda$ = 5.2~µm~\cite{REF20_Karnik}, corresponding to $\alpha$ = 1.17 dB/cm. This was a further motivation to design and measure racetrack resonators and test them at both $\lambda$~=~8.5~µm and $\lambda$~=~4.6~µm in our semiconductor platform.

\section{High Quality factor racetrack resonators}
As mentioned above, an alternative way to generate frequency combs involves pumping ultra-high-Q optical resonators with embedded Kerr nonlinearities using a continuous-wave laser. An efficient way to model these systems relies on the Lugiato-Lefever equation (LLE)~\cite{REF21_Lugiato}, that is a form of non-linear Schrödinger equation. This approach models the dynamics of the total intracavity field, and permits to infer a simple, closed-form formula for the minimum power needed to achieve parametric gain, as developed in~\cite{REF10_Chembo}:

\begin{equation}
P_{min} = 2\pi \alpha \frac{\omega_{L}^{2}}{8\gamma v_{g}^{2}}\frac{Q_{ext,t}}{Q_{tot}^{3}} = 2\pi \alpha \frac{\omega_{L}^{2}}{8v_{g}^{2}}\frac{cA_{eff}}{\omega_{0}n_{2}}\frac{Q_{ext,t}}{Q_{tot}^{3}}
\label{eq:Eq4}
\end{equation}

Where \textit{a} is the resonator radius; $\omega_{L}$ is the input laser frequency; $\omega_{0}$ is the cavity mode frequency; $Q_{ext}$ is the coupling Q-factor; $Q_{tot}$ is the total Q-factor; $A_{eff}$ is the effective surface of the waveguided mode; $n_{2}$ is the Kerr non-linear index. While a numerical study of the LLE is necessary to optimize a micro-comb device, this formula permits to rapidly grasp the crucial parameters that need to be considered: (i) Q factor, as P$_{\textnormal{min}}$ scales as 1/$Q^2$; (ii) effective volume ($A_{eff} \cdot 2\pi a$); and (iii) non-linear index. We have therefore implemented racetrack resonators not only to precisely measure the waveguide losses, but also to gauge the potential of our material platform for mid-IR frequency comb generation.

\subsection{Design Strategies}
\begin{figure}[H]
  \centering
  \includegraphics[width=0.7\textwidth]{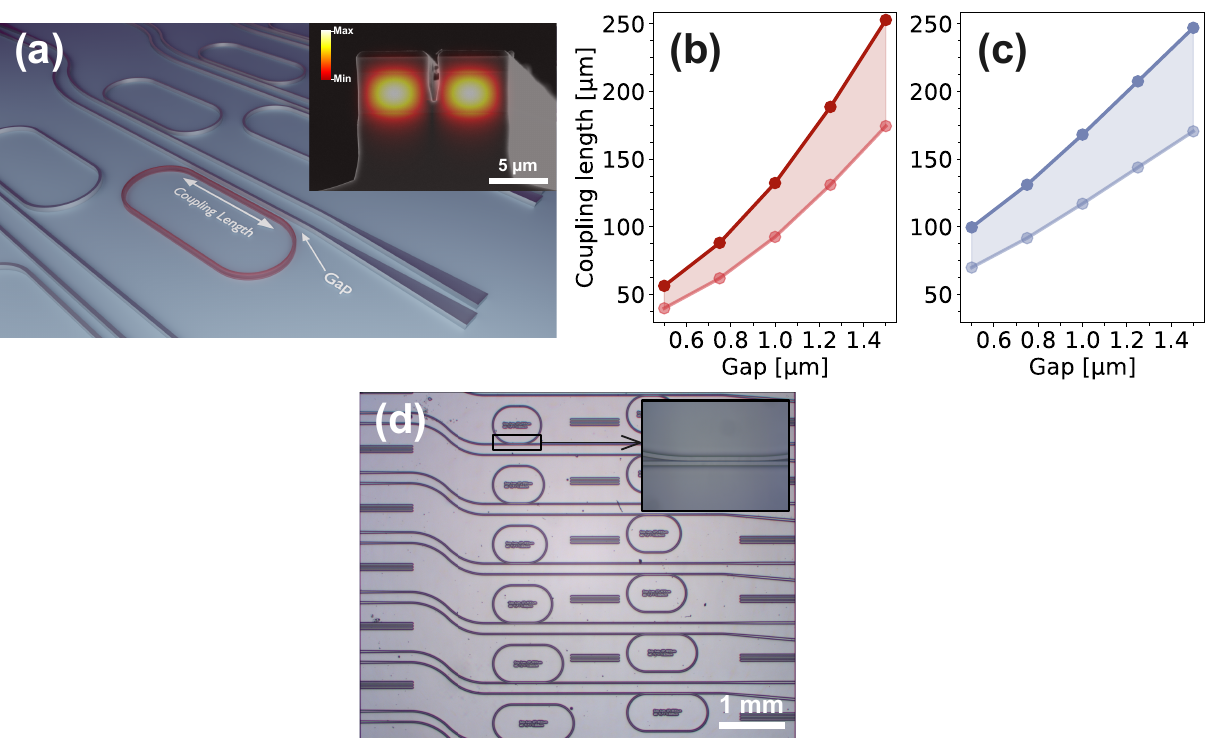} 
  \caption{\textbf{Design and geometry of the racetrack resonators.} \textbf{(a)} Schematics of the resonators. The inset shows a cross-section SEM image of the region between the racetrack and the bus waveguide, overlaid with the TM super-mode electric field profile. \textbf{(b)} Calculated critical coupling lengths as function of the gap between the racetrack and the bus waveguide for the TM fundamental mode. Two loss values are considered: 1.87 dB/cm and 0.87 dB/cm. \textbf{(c)} Same as \textbf{(b)}, but for the TE fundamental mode. \textbf{(d)} Optical microscope image (top view) of the chip with resonators of varying sizes and coupling lengths. The inset is a close-up of the coupling region.}
  \label{fig:Fig3}
\end{figure}

Racetrack resonators were designed and optimized to operate at $\lambda$~=~8.5~µm. The bus and the resonator waveguide ridge width was set to 6~µm, similarly to the spirals described in Section~\ref{sec:Section_2a}. FDTD simulations were performed to estimate the minimum bending radius that would make radiation and mode mismatch losses in the bent sections negligible. We found a minimum radius of 300~µm, hence, we set it to 350~µm. The design of the coupler (Fig.~\ref{fig:Fig3}a) was performed combining coupled mode theory (CMT)~\cite{REF22_Bogaerts} and FDE simulations, where gaps and coupling lengths L$_{\textnormal{c}}$ were tailored to achieve different coupling regimes. To assess the impact of the platform losses on the estimated L$_{\textnormal{c}}$, a range of propagation losses between 1.87~dB/cm and 0.87~dB/cm was considered. To improve accuracy, the calculation is performed assuming a 4.8-µm-deep etch in the gap, simulating the profile obtained from previous etching recipe optimizations. The final result is presented in Fig. \ref{fig:Fig3}b where the discrepancy in L$_{\textnormal{c}}$ between TM and TE polarizations is ascribed to stronger mode coupling in TM because of a lower confinement factor with respect to TE polarization. Coupling lengths in the [100 – 1000]~µm range were selected for the final fabrication, to span all the coupling regimes (critical-, under- and over-coupling). The gap is fixed to 1~µm due to limitations imposed by the fabrication process. The bus waveguide geometry is the same as the one in Section~\ref{sec:Section_2b}. However, the output facet of odd-numbered waveguides is tilted by 5$^\circ$: this reduces the reflectivity 
and reduces the amplitude of the Fabry-Pérot oscillations, simplifying the analysis of the racetrack resonators resonances.\\

The inset of Fig. \ref{fig:Fig3}a displays an SEM image of the coupling section right before the hard mask removal step. The etching penetrates only partially in the gap, leaving 30$\%$ of the core layer un-etched. This yields a stronger mode coupling, thus modifying the critical coupling condition (L$_{\textnormal{c}}$ should be shorter). A top view image of the sample is shown in Fig. \ref{fig:Fig3}d.

\subsection{Q factors at 4.6~µm and 8.5~µm wavelengths}
\begin{figure}[h]
  \centering
  \includegraphics[width=0.7\textwidth]{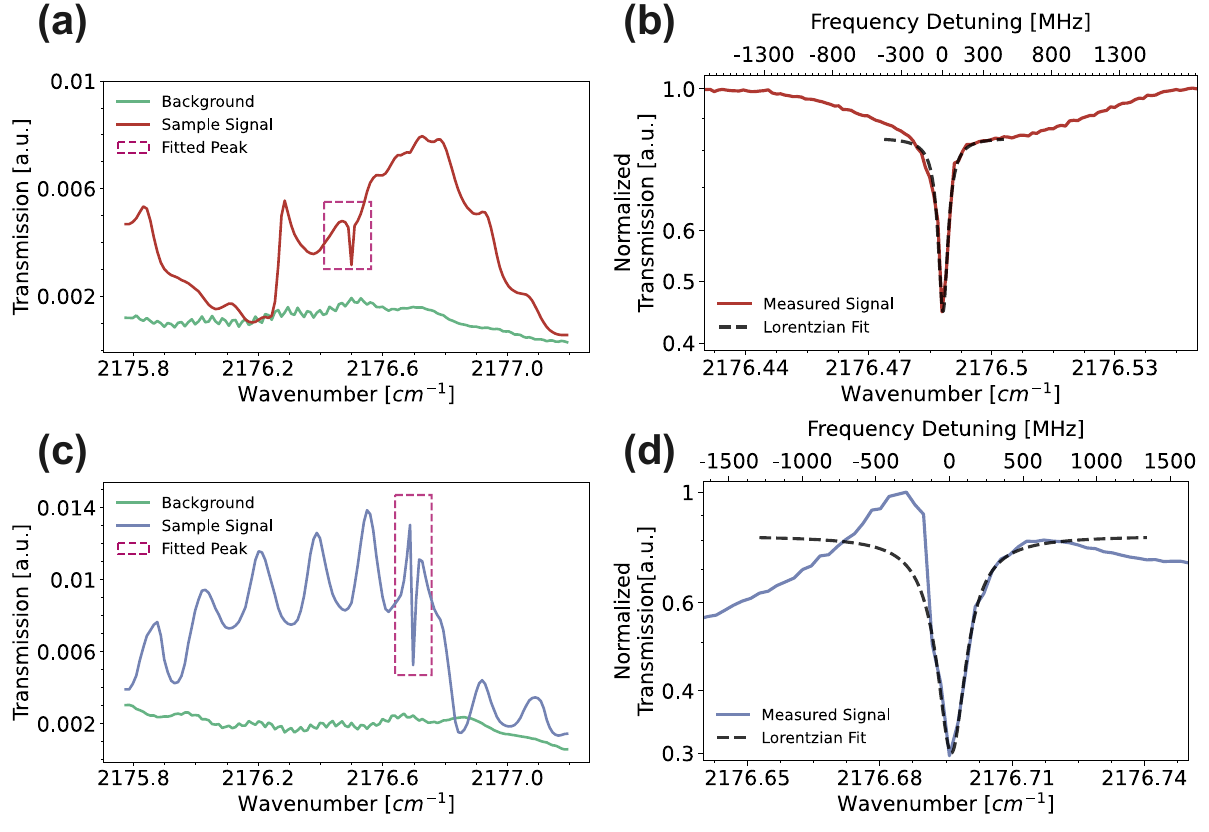} 
  \caption{\textbf{Experimental transmission spectra at $\lambda$~=~4.6~µm – via the the bus waveguide – of a racetrack resonator with L$_{\textnormal{c}}$ = 100~µm.} \textbf{(a)} Transmission spectrum (red curve) in TM polarization, showing a single resonance within the measurement span (limited by the DFB laser tuning range). The green curve represents the reference signal, measured without the sample. \textbf{(b)} Close-up of the resonance highlighted by the dashed rectangular region in panel \textbf{(a)}. The normalized transmission spectrum (signal/reference) is shown. The black dashed curve is a Lorentzian fit, yielding a record-high $Q_{loaded}$ of 6.4 $\times$ 10$^5$ and an estimated Qintrinsic of 7.4 $\times$ 10$^5$, assuming an undercoupling operating regime. The top horizontal axis corresponds to the frequency detuning with respect to resonance center. \textbf{(c)}, \textbf{(d)} Same as \textbf{(a)} and \textbf{(b)} but in TE polarization. Measured Qloaded is 2.1 $\times$ 10$^5$, with an estimated $Q_{intrinsic}$ of 2.4 $\times$ 10$^5$.}
  \label{fig:Fig4}
\end{figure}

Figure \ref{fig:Fig4} reports the measurements of the loaded Q-factors of the racetrack resonators at 4.6~µm wavelength, obtained via transmission measurements of waveguides coupled to racetrack resonators, in TM (upper panels) and TE (lower panels) polarizations. The measurements were performed as above, by spectrally tuning the emission of a commercial DFB-QC laser with the injection current. The current-wavelength correspondence was initially calibrated using a Fourier transform infrared (FTIR) spectrometers. Panel (a) shows a relatively large spectral scan: sample and background signals are reported, and a narrow resonance (circled in dashed) is evident. Panel (b) is a high-resolution close-up of the resonance: the FWHM is only 101 MHz, that translates in a loaded Q-factor ($Q_{tot}$) of 6.4 $\times$ 10$^5$.\\

Using the standard formula $Q_{intrinsic} = \left(  2\cdot Q_{tot}\right)/\left(  1 \pm \sqrt{T} \right)$ , where T is the transmission at resonance (and assuming under-coupling operation), we obtain an intrinsic Q-factor ($Q_{intrinsic}$) of 7.4 $\times$ 10$^5$. This is a record-high value for ultra-high Q-factor resonators in a III-V semiconductor platform and – furthermore – it corresponds to waveguide losses of 0.24 dB/cm, in full agreement with the measurements performed with the Fabry-Perot method (see Fig. \ref{fig:Fig2}).

Panels (c) and (d) report the same measurements, albeit in TE polarization. We obtain a Q$_{load}$ of 2.1 $\times$ 10$^5$, with an estimated Q$_{intrinsic}$ of 2.4 $\times$ 10$^5$. These lower Q-factors could stem from the relatively shallow etch of the gap between access waveguide and racetrack resonator that add scattering losses  (Fig. \ref{fig:Fig3}a, inset). Because of the multimode operation  at such a short wavelength, this effect may be more pronounced for the TE polarization due to the higher overlap with the waveguide sidewalls. \\
While the core result of the work is the record-high Q-factor at $\lambda$~=~4.6~µm, we also performed Q-factor measurements at 8.5~µm wavelength. Figure 5 shows the normalized transmissions, in TM and TE polarizations, around a racetrack resonator resonance. 

\begin{figure}[H]
  \centering
  \includegraphics[width=0.8\textwidth]{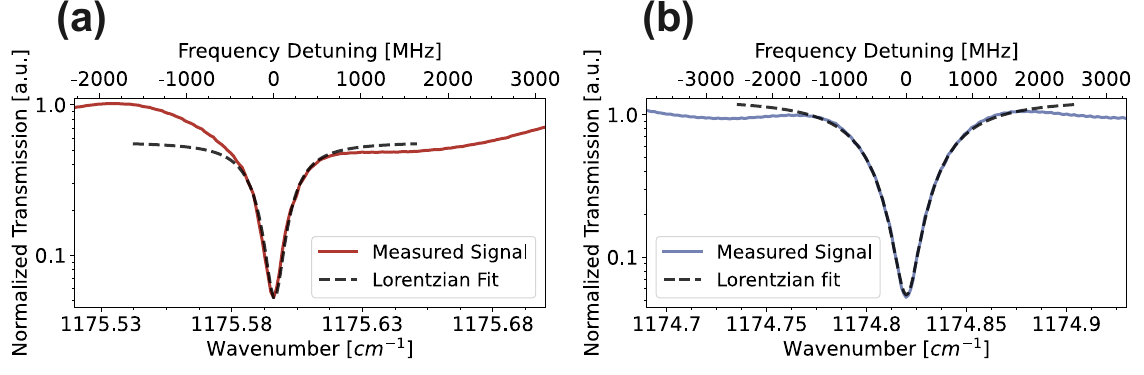} 
  \caption{\textbf{Experimental normalized transmission at $\lambda$~=~8.5~µm of the racetrack resonators.} \textbf{(a)} Normalized transmission of a TM resonance (close-up view), with a Lorentzian fit (black dashed curve) yielding $Q{loaded}$~= 5.8 $\times$ 10$^4$ and $Q_{intrinsic}$ = 8.5 $\times$ 10$^4$, assuming undercoupling operating regime. \textbf{(b)} Normalized transmission of a TE resonance (close-up view) for a different racetrack resonator, yielding $Q_{loaded}$~= 2.2 $\times$ 10$^4$ and $Q_{intrinsic}$ = 5.4 $\times$ 10$^4$, assuming overcoupling operating regime.}
  \label{fig:Fig5}
\end{figure}

Intrinsic Q-factors of 85.000 (54.000) in TM (TE) polarization, respectively, are observed. These lower values with respect to what is obtained at shorter wavelengths are in agreement with previous findings in the literature~\cite{REF15_Zhang}~\cite{REF23_Koompai}: the waveguide losses dramatically increase moving from the 4 µm to the 9 µm region. This result also suggests that our devices are most probably limited by free-carrier absorption, leaving ample opportunities for further Q-factor improvement by working on the material quality. For instance, the background doping – and therefore the propagation losses - could be further reduced by iron-doping the semiconductor material during the epitaxy~\cite{REF15_Zhang}~\cite{REF16_MontesinosBallester2024}.

\section{Conclusions/Perspectives}
We have demonstrated mid-IR integrated race-track resonators in the InGaAs-on-InP semiconductor platform with measured quality factors larger than 600.000 at a wavelength of 4.6~µm. Additionally, we have performed a full characterization of the optical propagation losses in this system at two representative wavelengths for the 1$^{\textnormal{st}}$ and 2$^{\textnormal{nd}}$ atmospheric transparency windows (4.6~µm and 8.5~µm).
These results open realistic vistas for the development of integrated mid-IR devices based on InGaAs-on-InP where the onset of stimulated parametric processes could be reachable, with the long-term goal of demonstrating Kerr micro-combs in the first atmospheric window (3-5 µm). The next crucial steps to achieve this ambitious goal are (i) achieving Q-factors beyond one million, and (ii) carefully engineering the waveguide dispersion in the proper parameter range for soliton generation. Plugging these values into Eq.\ref{eq:Eq4} yields parametric threshold values below 50 mW intracavity power, that are achievable for instance using commercial CW DFB QC lasers.

\vskip 1cm

\subsection*{Acknowledgements}
This work was partially supported by the French National Research Agency: project BIRD (No. ANR-21-CE24-0013) and from the European Union through FET-Open Grant NEHO (No. 101046329).             
This work was performed within the C2N micro nanotechnologies platforms and partly supported by the RENATECH network and the General Council of Essonne.
We thank M. Jeannin for technical help and D. Morini for useful discussions.


\makeatletter
\renewcommand{\refname}{}        
\renewcommand{\@biblabel}[1]{[S#1]}
\makeatother

\vspace{+1ex} 


\makeatletter
\renewcommand{\refname}{References}
\renewcommand{\@biblabel}[1]{[#1]}
\makeatother

\end{document}